\documentclass[twocolumn,prl,floatfix,showpacs]{revtex4}
\usepackage{graphicx}

\newcommand{\be}{\begin{equation}}
\newcommand{\ee}{\end{equation}}
\newcommand{\barr}{\begin{eqnarray}}
\newcommand{\earr}{\end{eqnarray}}
\newcommand{\breakeq}{\nonumber \\ &&}
\newcommand{\unit}[1]{\mathrm{\,\,#1}}
\newcommand{\diel}{\epsilon_0}

\begin{document}

\title{Two-Electron Photon Emission from Metallic Quantum Wells}
\author{Germar Hoffmann and Richard Berndt}
\affiliation{Institut f\"ur Experimentelle und Angewandte Physik,
        Christian-Albrechts-Universit\"at zu Kiel, D--24098 Kiel, Germany}
\author{Peter Johansson}
\affiliation{Department of Natural Sciences,  University of \"Orebro,
        S--701\,82 \"Orebro, Sweden}

\date{\today}

\begin{abstract}
Unusual emission of visible light is observed in scanning tunneling
microscopy of the quantum well system Na on Cu(111). Photons are emitted at
energies exceeding the energy of the tunneling electrons. Model calculations
of two-electron processes which lead to quantum well transitions reproduce
the experimental fluorescence spectra, the quantum yield, and the power-law
variation of the intensity with the excitation current.
\end{abstract}


\maketitle


Tunneling electrons in a scanning tunneling microscope (STM) can excite
vibrational or electronic modes of the sample by  inelastic tunneling
provided that their energy exceeds the excitation energy. These
excitations have been detected either by their contribution to the
tunneling current \cite{HO} or by investigating light that is being emitted
from the tunneling gap \cite{Jim,NAPRL}. Thus, inelastic tunneling
spectroscopies have been performed. Usually, tunneling electrons can safely
be assumed to be independent of each other in these spectroscopies. Even at
a high tunneling current $I = 100 $\ nA the average time between two
consecutive tunneling events is $\sim 1.6$\ ps. Assuming Poisson
statistics, two electrons are rather unlikely to interact in the tunneling
gap. As a consequence, inelastic processes which involve multiple electrons
have only been observed in the particular case of STM-induced desorption of
H from Si.  The lifetime of the H-Si stretch mode which is involved in
desorption is in the range of nanoseconds \cite{fol:98} enabling an
interaction with several consecutive electrons before deexcitation.

Here, we report on unusual emission of visible light from Na on Cu(111),
a metallic system which exhibits well-studied quantum well states (QWS)
near the Fermi energy $E_F$
\cite{wal:80,lin:87,fis:91,dud:91,fis:94,car:97,car:00,kli:01}.
Surprisingly, fluorescence spectra reveal the emission of ``forbidden''
photons whose energy $h \nu$ significantly exceeds the energy of a
tunneling electron $eU$, where $U$ is the sample voltage. The intensity
of the ``forbidden'' light increases approximately like $I^{1.5}$ where
$I$ is the tunneling current, with the exponent decreasing to 1.2 at the
highest currents used. Its quantum efficiency reaches values of up to
$\sim 10^{-7}$ photons per tunneling electron at large $I$.

Electronic lifetimes at the Na/Cu(111) surface being on a fs timescale
\cite{car:97} we are lead to conclude that two-electron processes are
involved which do not rely on a stepwise accumulation of energy in an
excited mode. We propose a model where two electrons tunnel more or less
simultaneously.  Once they are in the vacuum-barrier region between tip
and sample they may exchange energy through the Coulomb interaction
which is relatively unscreened there. As a result of this Auger-like
process, one of the electrons can emit a photon with $h \nu > eU$.
Despite the simplicity of the model, calculated fluorescence spectra and
the current dependence of the quantum efficiencies are comparable to the
experimental data.

Spectral structure extending beyond the condition $h \nu < e U$ has been
reported for photon emission from Au films investigated at ambient
temperature. However, no explanation of this intriguing result is currently
available \cite{BEAU}. Uehara et al.\ \cite{UE} reported on light emission
at $h \nu = 2 \, e U$ from  superconducting Nb tips and samples
at $T = 4.7$ K and explained this emission in terms of Cooper-pair
tunneling.  Photon emission at large $h \nu$ has also be observed from
metal point contacts which emit black-body radiation at elevated
currents \cite{DOW}.

Our experiments were performed with a ultra-high vacuum (UHV) STM
operated at a temperature $T=4.6$~K \cite{DrKli}. W tips were prepared
by electrochemical etching and subsequent sputtering and annealing in
UHV. The Cu(111) surface was cleaned by repeated cycles of Ar-ion
bombardment and annealing. Na films were evaporated from outgassed SAES
Getters sources onto the Cu crystal held at room temperature. Na
coverages were calibrated using the binding energies of the lowest QWS
\cite{car:97}. After preparation at room temperature the samples were
transferred to the STM and cooled to $T=4.6$~K\@. Photons in the energy
range 1.1 eV $< $ h$\nu < 3.5$ eV were detected with a lens-system in
UHV, coupling the light to a grating spectrometer and a liquid nitrogen
cooled CCD camera \cite{hof02}. The spectra have been corrected for the
wavelength dependency of the detection efficiency. For the voltages used
here, up to currents of $\sim 100$ nA, surface modifications was rarely
observed on flat surfaces.  While surface modification becomes more
probable at higher currents, during acquisition of the data sets shown
here no such events occured as verfied from STM images and simultaneous
monitoring of the vertical tip position.

\begin{figure}[htb]
\includegraphics[angle=0, width=8cm]{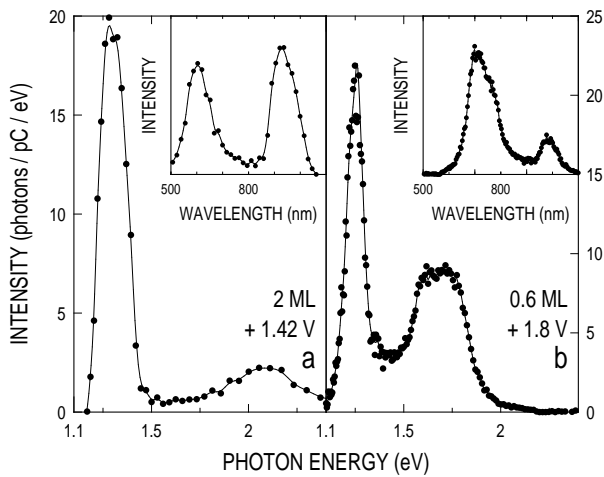}
\caption{Fluorescence spectra from Na monolayers on Cu(111). (a) 2
ML Na, $U=1.42$ V, $I=100$ nA\@. (b) 0.6 ML Na, $U=1.80$ V,
$I=357$ nA\@. Solid lines serve to guide the eye. The data have
been corrected for detector response. However, due uncertainties
of the rapidly decreasing detector sensitivity at photon energies
below $\sim 1.25$ eV the correction is reliable only at higher
photon energies. The insets show the uncorrected data as measured
(in counts vs.\ wavelength in nm) } \label{spex}
\end{figure}

Figure 1 displays fluorescence spectra recorded at elevated
tunneling currents from a (a) 2 ML and (b) 0.6 ML Na film on
Cu(111). The QWS binding energies are known from tunneling
spectroscopy \cite{SHIFT}.  At 2 ML, unoccupied states exist at
$E_1 = 0.15$ eV and $E_2 = 2.2$ eV\@. At 0.6 ML, these states are
located at $E_1 = 0.4$ eV and $E_2 = 2.1$ eV\@. Previously, photon
emission due to two processes has been reported from these layers
\cite{NAPRL,NaPRB}. At low $U$, i.\,e. $eU<E_2$, electrons tunnel
inelastically from the tip Fermi level to the lower QWS and emit
photons. Fluorescence spectra reveal a maximum which shifts with
$eU$. When $eU > E_2$, tunneling to the upper QWS occurs and a
subsequent transfer of an electron to the lower QWS gives rise to
the emission of quantum well luminescence at $h \nu = E_2-E_1$.
Enhancement by a local plasmon renders these processes efficient.
The data of Fig.\ 1 appears to be consistent with this picture.
Two spectral components are discernible, an emission at low photon
energies involving inelastic tunneling and an additional peak at
(a) $h \nu \sim 2.0$ eV  and (b) $h \nu \sim 1.7$ eV which is due
to transitions between QWS.  What is new in Fig.\ 1 is the fact
that these data were recorded at a sample voltage (a) $U=1.42$ V
and (b) $U=1.80$ V\@. In Fig.\ 1a the entire quantum well emission
peak seems to violate energy conservation, $h \nu \sim 2$ eV $>
eU$ with a maximum energy excess of $\sim 0.7$ eV\@.
In Fig.\ 1b there is still significant intensity with $h \nu
> eU$. However, the quantum well emission at $h \nu \sim 1.7$ eV
in Fig.\ 1b becomes even more surprising if one recalls that the
upper QWS, which is involved in the underlying transition, is
located at $E_2=2.1$ eV which is substantially larger than $eU$.

\begin{figure}[htb]
\includegraphics[angle=0, width=8 cm]{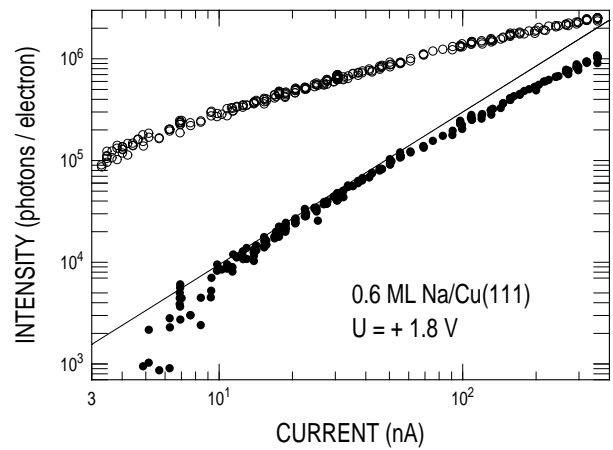}
\caption{Intensity of quantum well emission (dots, 1.45 eV $< h
\nu < 2$ eV, corrected for detection efficiency) and emission due
to inelastic tunneling (circles, 1.1 eV $< h \nu < 1.45$ eV) vs.\
current $I$ evaluated from 830 fluorescence spectra of a 0.6 ML
coverage at $U=1.8$ V\@. As a guide to the eye a slope of
$I^{1.5}$ is indicated (line).} \label{log-log}
\end{figure}

Two-electron processes provide a natural explanation of the
unusual emission in Fig.\ 1. Since such processes imply a
nonlinear variation of the intensity with the tunneling current
$I$ we recorded series of some 800 fluorescence spectra while
varying $I$ and evaluated the ``forbidden'' intensity. The
double-logarithmic plot in Fig.\ 2 reveals that the intensity
scales approximately like $I^{1.5}$ confirming the above
explanation.  As a consequence, the ``forbidden'' emission is weak
at low tunneling currents.  That is why it has been overlooked
previously. In addition to its variation with $I$, the intensity
of the quantum well emission from 0.6 ML Na also depends strongly
on the voltage $U$ for $U<2$ V\@.  Consequently, the quantum
efficiency of the ``forbidden'' emission varies significantly
depending on the specific $I$ and $U$ chosen.  We estimate an
efficiency on the order of $10^{-7}$ photons per tunneling
electron at $I=100$ nA and $U=1.8$ V from 0.6 ML Na.

A possible explanation of the observed two-electron processes appears to be
tunneling of an electron into a long-lived empty state of the Na/Cu(111)
surface and subsequent further excitation of this electron via interaction
with a second tunneling electron.  We estimated the probability of such
processes assuming a Poisson distribution of the intervals between
tunneling events.  To obtain the observed quantum
efficiencies an excited electronic state with a lifetime $\tau \approx 1$
ps needs to be postulated.  This value is much larger than typical
electronic lifetimes at surfaces.  Moreover, it is unclear why an electron
should remain localized under the tip over this extended period of time. We
therefore discard this type of mechanism.

We have instead considered two other two-electron mechanisms that we
believe cause emission of photons with an energy exceeding $eU$: (i) A
coherent Auger-like process in which energy is transferred from one
tunneling electron to another. (ii) Decay of the hot holes \cite{Gadzuk}
that are injected into the tip because most of the tunneling current passes
through the lower QWS. The decaying holes create hot electrons in the tip
which subsequently can tunnel into the upper QWS and thereby cause photon
emission.

The Keldysh Green's function (GF) formalism provides a
suitable theoretical framework for calculating the intensity of the
emitted light from a system out of equilibrium
such as an STM under finite bias.
The intensity can be written \cite{NaPRB}
\begin{equation}
   \frac{dP}{d\Omega d(\hbar\omega)}
   =
   \frac{\omega^2 |G(\omega)|^2}{16\pi^3\diel c^3 \hbar}
   \int_{V} d^3r
   \int_{V} d^3r'
   \, i \, \Pi^{<}(\vec{r}, \vec{r}', \omega).
\label{keldysh1}
\end{equation}
The integrations run over a volume between the tip and sample where the
electrons and photons interact efficiently. $G$ is a factor describing the
enhancement of the electromagnetic vacuum fluctuations in the tip-sample
cavity, and $\Pi^{<}$ is the Fourier transform of the current-current GF,
$-i\langle j_z(\vec{r}', 0) j_z(\vec{r}, t)\rangle$, which in the case of
{\em allowed} light emission can be expressed in terms of a current matrix
element between the initial and final electron states \cite{NaPRB}. The
detailed calculations of electron states and matrix elements employs a
one-dimensional model for the system.  The Cu potential is corrugated to
yield a band gap of 5 eV at the Brillouin zone boundary,  while the
potential in the Na layer and the tip are assumed to be constant \cite{LW}.
A tilted square barrier, rounded and lowered by image-potential
contributions, separates the electrodes.  In addition, the potential in the
Na layer is given an imaginary part $- i \Gamma$, with $\Gamma=$ 0.1 eV, to
mimic the electron scattering processes that limit the lifetime of the
quantum well states.

For {\em forbidden} light emission through an Auger-like process, the
leading contribution to the integrals in Eq.\ (\ref{keldysh1}) can be
written in terms of a sum of the squares of second-order matrix elements,
\barr
   &&
   \frac{dP}{d\Omega d(\hbar\omega)}
   =
   \frac{\omega^2 |G(\omega)|^2}{8\pi^2\diel c^3}
   \,
   \sum_{k_1 k_2 q}
   |M_{k_1,k_2,q}|^2
   \breakeq
   \times
   \delta(E_{k_1} + E_{k_2} - E_{1,k_1+q} -E_{1,k_2-q} - \hbar\omega).
\earr
$M_{k_1,k_2,q}$ describes how two electrons in the tip labeled by the
momenta $\vec{k_1}$ and $\vec{k_2}$ first interact through a screened
Coulomb interaction, $e^2
e^{-\kappa|\vec{r_1}-\vec{r_2}|}/(4\pi\diel|\vec{r_1}-\vec{r_2}|)$ and
exchange energy and momentum\cite{kappanote}. One electron goes into the
lower QWS [with in-plane momentum $\vec{k_{1,\|}}+\vec{q}$ and energy
$E_{1,k_1+q}= E_1 +\hbar^2(\vec{k_{1,\|}}+\vec{q})^2/(2m)$] directly and
takes no further part in the process, while the other eventually emits a
photon in a transition from an intermediate state into  the lower QWS (with
in-plane momentum $\vec{k_{2,\|}}-\vec{q}$ and energy $E_{1,k_2-q}$). When
the two electrons have opposite spin we can write
\barr
   &&
   M_{k_1,k_2,q}
   =
   \frac{-i e\hbar}{2m}
   \int_{V} d^3r
   \int_{V} d^3r_1
   \int_{V} d^3r_2
   \breakeq
   \times
   \left\{
      \phi_{k_2-q}^{*}(\vec{r})
      \,
      \frac{\partial g^{r}}{\partial z} (\vec{r}, \vec{r_2})
      -
      \frac{\partial \phi_{k_2-q}^{*}}{\partial z}
      \,
      g^{r} (\vec{r}, \vec{r_2})
   \right\}
   \breakeq
   \times
   \phi_{k_1+q}^{*}(\vec{r_1})
   \,
   \frac{e^2\, e^{-\kappa|\vec{r_1}-\vec{r_2}|}}
   {4\pi\diel |\vec{r_1} - \vec{r_2}|}
   \,
   \psi_{k_2}(\vec{r_2})
   \,
   \psi_{k_1}(\vec{r_1}),
\earr
where $\phi$ denotes the QWS wave function, while $\psi$ stands for tip
wave functions. The retarded electron Green's function $g^r$ describes the
propagation of the electron in the intermediate state before photon
emission. For energy and momentum conservation to hold in the photon
emission process the energy and in-plane momentum in the intermediate state
must be $E_{1,k_2-q}+ \hbar\omega$ and $\vec{k_{2,\|}}-\vec{q}$,
respectively. The electron Green's function has a resonance when its energy
argument coincides with the energy of the upper QWS which explains why, as
we will see, the forbidden light emission mainly produces photons with
energy $h\nu = E_2-E_1$.

We have also calculated the light emission intensity as a result of
hot-hole decay. To this end we studied a semiclassical model based on Ref.\
\onlinecite{Ritchie} for hot-hole-electron cascade and diffusion (with an
elastic mean free path of 2 nm) in the tip, and calculated the influx of
secondary hot electrons onto the tip apex. This influx was then used as
input in a calculation of the light emission intensity along the line of
Ref.\ \onlinecite{NaPRB}.

\begin{figure}[htb]
\includegraphics[angle=0, width=7 cm]{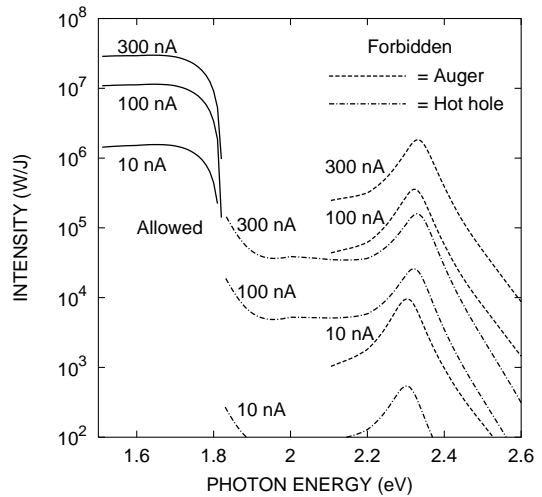}
\caption{``Allowed'' and ``forbidden'' light emission calculated for a
model system with QWS at $E_1$\,=\,0.2 eV and $E_2$\,=\,2.5 eV\@. Results
are shown for various currents indicated in the figure and a voltage
$U$\,=\,2 V\@. The thickness of the Na overlayer was set to 0.613 nm,
corresponding to 2 monolayers. }
\label{calc}
\end{figure}

Figure 3 displays results from our model calculations obtained for $U$\,=\,2
V and $I$\,=\,10, 100, and 300 nA, respectively. The model potential leads
to $E_1$\,=\,0.2 eV and $E_2$\,=\,2.5 eV at $U$\,=\,2 V\@. Under these
conditions, the upper quantum well state at $E_2$ is not accessible.
Moreover, the energy $eU$ of  a single electron is not sufficient for
exciting the corresponding quantum well transition. As a result,
one-electron processes (solid lines in Fig.\ 3) give rise to plasmon
mediated emission by inelastic tunneling from the tip Fermi level only to
the lower QWS.  The emission occurs predominantly at $h \nu < e U - E_1$ as
expected \cite{NAPRL,NaPRB}.

The emission calculated for the Auger-like and hot-hole processes,
respectively, are indicated by dashed lines. We do find sizable
emission, which is about one order of magnitude stronger for the Auger
process than the hot-hole mechanism, peaked at $h \nu \sim 2.3 $ eV
(thus  $h \nu>eU)$ due to quantum well transitions.  The electron that
eventually causes the light emission gains enough energy (i.e.\ $\approx
0.5 \unit{eV}$), either through Coulomb interactions with another
electron while tunneling, or in the hot-hole-electron cascade in the
tip, to be promoted to the upper quantum well resonance situated above
the tip Fermi level in energy. We note that the particular electronic
structure of Na on Cu(111) with states at well-defined energies is
essential in achieving significant signal levels.

The calculations predict quantum yields of up to $10^{-7}$ photons per
electron for the Auger-like mechanism at $I$=100 nA and $U$=2.3 V in
reasonable agreement with the experimental value. Given the experimental
uncertainty of absolute photon intensities, as well as the
approximations involved in the calculations, a comparison of its
variation with $I$ is more significant. The Auger-like process yields
$I^{1.5}$ close to the experimental data \cite{TOKI}. The hot-hole
process gives a slightly larger exponent, 1.6.   While both mechanisms
must be considered as plausible explanations for the forbidden light
emission the larger calculated intensities  indicate that the Auger
process is the dominating one.

In summary, we reported on unusual STM-induced photon emission from a
metallic quantum well system at photon energies exceeding the limit $h \nu
\leq e U$. Model calculations revealed that owing to the particular
electronic structure of Na on Cu(111) two-electron processes can cause
quantum well transitions and corresponding fluorescence.  Similar effects
may be observable in other quantum confined systems.

We acknowledge support by the European Commission (TMR {\it EMIT}),  the
Deutsche Forschungs\-gemein\-schaft and the Swedish Research Council
(VR).


\begin{thebibliography}{0}
\bibitem{kappanote}
The phenomenological screening parameter $\kappa$ is introduced because
Coulomb interactions in the vacuum region are screened by image charges
induced in the nearby electrodes.  We have used the constant value
$\kappa=(1 \unit{nm})^{-1}$ consistent with typical experimental values
for the tip-sample separation.
\end{thebibliography}
\end{document}